\documentclass[reprint, amsmath,amssymb, aps, groupedaddress, a4paper]{revtex4-1}
\usepackage[a4paper,margin=2cm]{geometry}

\usepackage{amsmath}
\usepackage{graphicx}
\usepackage{dcolumn}
\usepackage{bm}
\usepackage[hidelinks]{hyperref}
\usepackage[mathlines]{lineno}
\usepackage{placeins}

\usepackage{xcolor}
\usepackage{siunitx}
\usepackage[version=4]{mhchem}

\begin{document}

\author{Joan~Sendra}
\affiliation{Laboratory for Nanometallurgy, Department of Materials, ETH Zurich, Zurich, Switzerland}

\author{Fabian~Haake}
\affiliation{Laboratory for Nanometallurgy, Department of Materials, ETH Zurich, Zurich, Switzerland}

\author{Micha~Calvo}
\affiliation{Laboratory for Nanometallurgy, Department of Materials, ETH Zurich, Zurich, Switzerland}

\author{Henning~Galinski}
\affiliation{Laboratory for Nanometallurgy, Department of Materials, ETH Zurich, Zurich, Switzerland}

\author{Ralph~Spolenak}
\email{ralph.spolenak@mat.ethz.ch}
\affiliation{Laboratory for Nanometallurgy, Department of Materials, ETH Zurich, Zurich, Switzerland}

\date{February 8, 2023}

\title[]{Scanning Reflectance Anisotropy Microscopy for Multi-Material Strain Mapping}

\keywords{phase-modulated microscopy, strain mapping, elasto-optic effect, metasurfaces, semiconductors, metals}

\begin{abstract}
Strain-engineering of materials encompasses significant elastic deformation and leads to breaking of the lattice symmetry and as a consequence to the emergence of optical anisotropy. However, the capability to image and map local strain fields by optical microscopy is currently limited to specific materials. Here, we introduce a broadband scanning reflectance anisotropy microscope as a phase-sensitive multi-material optical platform for strain mapping. The microscope produces hyperspectral images with diffraction-limited sub-micron resolution of the near-normal incidence ellipsometric response of the sample, which is related to elastic strain by means of the elasto-optic effect. We demonstrate cutting edge strain sensitivity using a variety of materials, such as metasurfaces, semiconductors and metals. The versatility of the method to study the breaking of the lattice symmetry by simple reflectance measurements opens up the possibility to carry out non-destructive mechanical characterization of multi-material components, such as wearable electronics and optical semiconductor devices.
\end{abstract}

\maketitle

\section{Introduction}

 In October of 1950, Bardeen and Shockley published their seminal work on deformation-potential theory that, for the first time, linked strain-induced shifts at the band edge to the strain tensor~\cite{bardeen_deformation_1950}. This fundamental work, soon validated experimentally for uniaxially stressed silicon~\cite{hensel_cyclotron_1963}, introduced a new field in solid state physics: strain-engineering~\cite{maiti_stress_2021}.
 \par
 Today, strain-engineering enables the enchancement of the optoelectronic properties of semiconductor materials, e.g. increasing photoluminescence~\cite{Tyurnina2019,suess_analysis_2013}, absorption~\cite{katiyar_breaking_2020}, photo~\cite{miao_strain_2022} and electrocatalysis~\cite{xia_strain_2019}, and carrier mobility~\cite{chu_strain_2009} among others. Furthermore, elastic strain is applied to improve magnetic~\cite{hu_enhanced_2020} properties or even raise the transition temperature of superconducting and ferroelectric materials~\cite{schlom_elastic_2014}.
 \par
 Strain-engineering, which also implies accurate strain control, is particularly important in the field of flexible and wearable electronics~\cite{Xu2017,kim_epidermal_2011}. These systems, often used in health care and monitoring devices, comprise a wide range of materials with very different mechanical properties, including polymers, metals and semiconductors, while having to perform under harsh mechanical deformation conditions~\cite{wang_skin-inspired_2018}. The requirement to withstand such conditions has made strain-engineering a critical point in the design process of flexible electronics, thus making mechanical and failure analysis of the utmost importance.
\par
However, current non-destructive strain analysis techniques require a lot of resources, such as time, funds and special equipment, and are often limited to a single class of materials. For example, synchrotron x-ray diffraction offers high resolution and strain sensitivity but requires beam time. Raman spectroscopy is a popular vibrational spectroscopic technique for standard semiconductors but is not suitable for strain measurements on metallic or Raman inactive samples.
\par
Here, we propose scanning reflectance anisotropy microscopy (SRAM) as an alternative tool for multi-material differential strain mapping. Classical reflectance anisotropy spectroscopy (RAS) is a near-normal incidence ellipsometric technique that measures the normalized difference of Fresnel coefficients along two orthogonal directions, able to directly probe anisotropy of the band structure at optical frequencies~\cite{volpi_sensing_2021, Aspnes1985,weightman_reflection_2005}. With its high sensitivity (reflectance differences down to $\Delta r/r \sim 10^{-5}$ can be measured) RAS is especially suited to detect breaking of the lattice symmetry due to strain~\cite{wyss_reflectance_2015,cole_stress-induced_2003,papadimitriou_highly_2005,Ortega-Gallegos2008} or crystal orientation. In particular, for the mechanical characterization of thin films, RAS has proved itself to be advantageous over x-ray diffraction due to its smaller probing volume and faster acquisition times~\cite{Wyss2017}.
\par

Despite the promising potential, classical RAS is fundamentally limited due to its low spatial resolution. Typical setups exhibit a probing spot of a couple of millimetres as increasing the angle of incidence for focusing poses problems due to the high polarization sensitivity of the technique~\cite{koopmans_microscopic_1998,shen_using_2017}. To compensate this short-comming, various setups have been proposed~\cite{Huang2016,zhang2005reflectance,lastras-martinez_microreflectance_2009,hu_rapid_2019,lastras-martinez_micro_2012} but the remarkable potential of this technique as versatile multi-material platform for strain mapping remains unexplored.

Here, we show how scanning reflectance anisotropy microscopy (SRAM) can serve as a robust and versatile platform for strain mapping in optically anisotropic media. Due to its high resolution, high signal-to-noise ratio and broad spectral bandwidth SRAM facilitates quantitative phase sensitive measurements at the diffraction limit. We discuss its design and demonstrate its multi-material capability performing microscopy on strained metasurfaces (metamaterials), strained suspended germanium bridges (semiconductors) and gold thin films (metals).
\begin{figure*}[ht!]
    \centering
    \includegraphics[width=\linewidth]{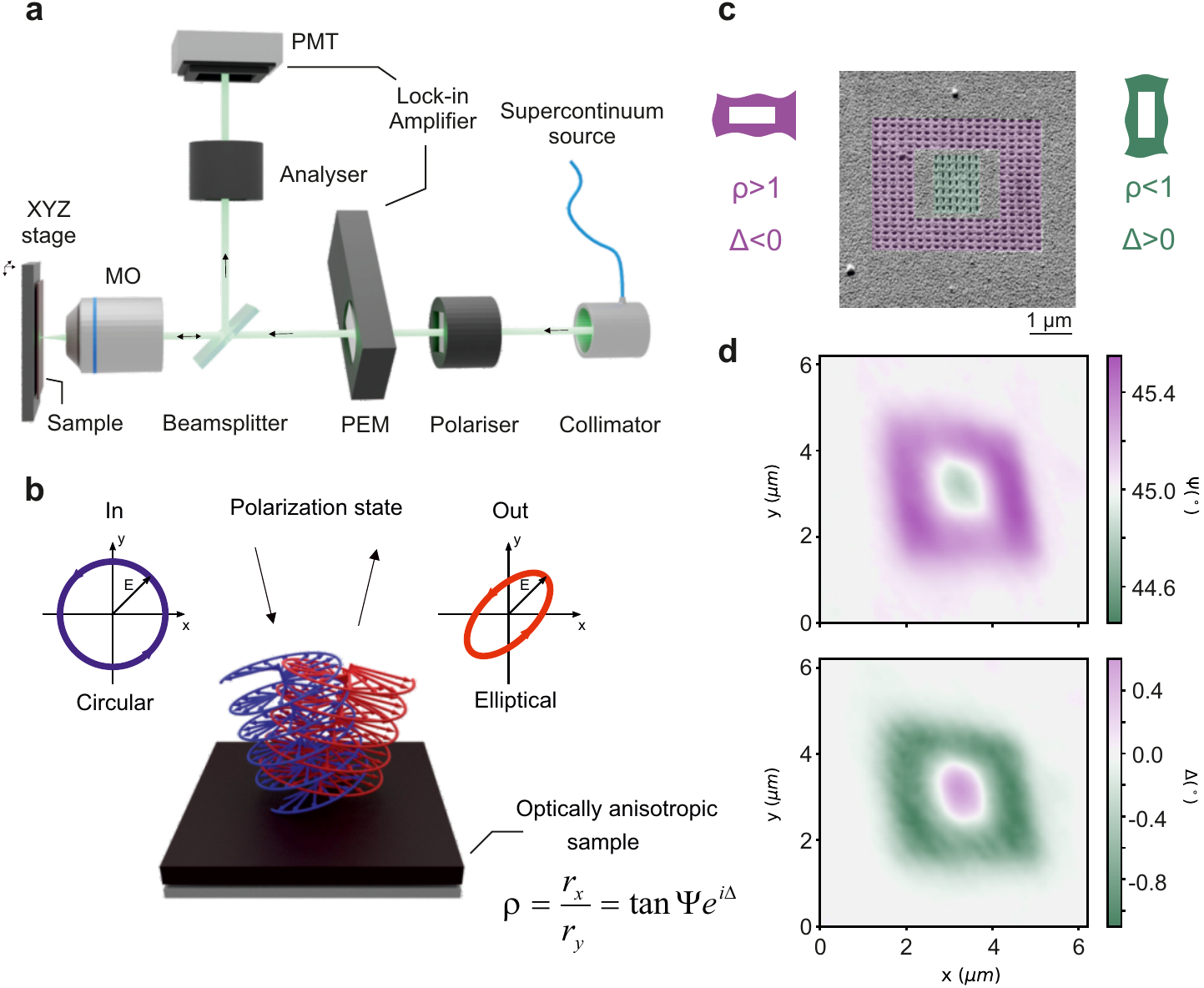}
    \caption{\textbf{a} Schematic of the scanning reflectance anisotropy microscope (SRAM)  displaying the laser beam path, including a photoelastic modulator (PEM), a microscopy objective (MO) and photomultiplier tube (PMT). \textbf{b} Illustration of the measurement principle of SRAM for a modulation state where the incident light is circularly polarized. In case of an optically anisotropic sample (in plane), the change in phase and amplitude of the out-going light can be represented by the sample's ellipsometric properties $(\Psi,\Delta)$ on the polarization state of the probing beam.  \textbf{c,d} SEM image and SRAM measurements of a metasurface composed of dipolar slot antennas showcasing the sensitivity of the setup to the ellipsometric properties (phase $\Delta$ and amplitude $\Psi$) of the sample. Phase sensitivity lower than 1$^{\circ}$ is achieved.}
    \label{fig1}
\end{figure*}
\section{Results}
\subsection{Principle of SRAM}

Scanning reflectance anisotropy microscopy (SRAM) uses phase modulation to infer the near-normal incidence ellipsometric response of the sample. Figure \ref{fig1}a shows the main optical components of the microscope and the light path. Using a transfer matrix analysis of the polarization elements in the setup~\cite{Martin2001} the intensity at the detector plane can be written as a Fourier series with coefficients being a function of the real and imaginary part of the SRAM signal,
\begin{widetext}
    \begin{equation}
        I\propto 1 + 2\Re\left(\Delta r/r\right)\sum_{m=0}^{\infty}J_{2m}(\Gamma_0)\cos(2m\omega t) + 2\Im\left(\Delta r/r\right)\sum_{m=0}^{\infty}J_{2m+1}(\Gamma_0)\sin\left[(2m+1)\omega t\right] ,
        \label{eq:1}
\end{equation}
\end{widetext}

where $J_n$ is the nth-order Bessel function, $\omega$ the operating frequency of the photoelastic modulator and $\Gamma_0$ the modulation amplitude chosen such that $J_0(\Gamma_0)=0$, and $\Delta r/r$ is given by $\Delta r/r=2\left(r_x-r_y\right)/(r_x+r_y)$. By using a lock-in amplifier (Figure \ref{fig1}a) one can extract $\Re\left(\Delta r/r\right)$ and $\Im\left(\Delta r/r\right)$ from the first and second harmonic components. While $\Re\left(\Delta r/r\right)$ is typically the value reported in classical RAS measurements, it does not provide an intuitive understanding of the sample's optical properties. Instead, from $\Delta r/r$ one can calculate the ellipsometric parameters,

\begin{equation}
    \rho=\frac{r_x}{r_y}=\frac{2-\Delta r/r}{2+\Delta r/r}=\tan\Psi e^{i\Delta} ,\label{rho}
\end{equation}

with $2\left(\Psi-\pi/4\right)\simeq\Re\left(\Delta r/r\right)$ and $\Delta\simeq\Im\left(\mathrm{\Delta} r/r\right)$ for small $\Delta r/r$~\cite{acher_reflectance_1992}. Here, $(\Psi,\Delta)$ provide a straightforward description of the effect the in-plane optical anisotropy of a material has on the reflected light (Figure \ref{fig1}b). Since SRAM is effectively a near-normal incidence technique, circumventing the angular dependence of the Fresnel coefficients, $\Psi$ and $\Delta$ directly quantify the anisotropic optical properties of the sample, making SRAM a phase-sensitive technique. More details are given in the supplementary information.
\par
To illustrate the phase sensitivity of the microscope, we designed an optically anisotropic metasurface based on nanoslot antennas that exhibits a dipolar resonance in the spectral range of the setup. Figure \ref{fig1}c depicts the metasurface containing an arrangement of nanoslot antennas with an outer-ring made of horizontally oriented antennas (width: 50 nm, height: 100 nm), and an array of vertically oriented antennas of the same size. The geometry of the nanoantennas translates to a polarization selectivity of the localized surface plasmons, thus giving rise to a resonance. Altering the nanoslot antennas' dimensions, pitch, and orientation allows us to engineer the ellipsometric parameters of the sample ($\Psi$,$\Delta$) as well as placing the resonance inside the spectral range of the microscope. Figure \ref{fig1}d shows the SRAM $\left(\Psi,\Delta\right)$ map of the nanoslot antenna arrangement at the resonance frequency of 2.39 eV,  displaying the symmetry breaking of the metasurface.  These measurements showcase the high phase sensitivity of the technique, able to distinguish phase differences smaller than 1$^\circ$ at diffraction-limited resolution.

\subsection{Microscopy of Strained Metasurfaces}

Given the high phase sensitivity, we investigate if SRAM can detect even small deformations of such a slot antenna. Uniaxial deformation of the antenna changes the ratio between the longitudinal and transversal axial lengths, thus changing the optical anisotropy of the antenna. This change is influenced by both the elongation along the strain axis and the resulting shortening along the orthogonal axis due to Poisson's effect. The change in optical anisotropy should result in a red-shift of the resonance frequency for positive strains parallel to the long axis of the antenna.
\par
\begin{figure*}[ht!]
    \centering
    \includegraphics[width=1\linewidth]{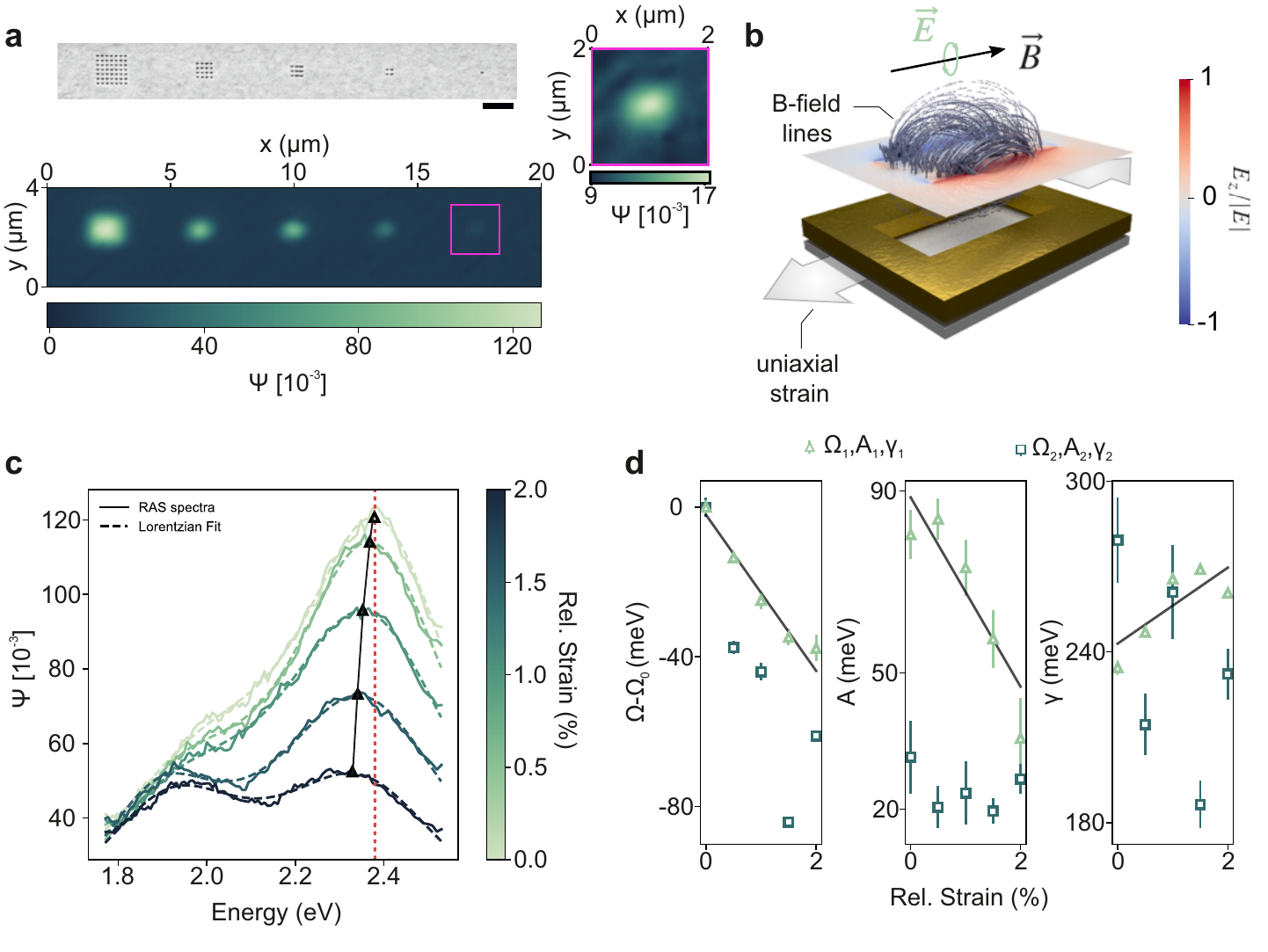}
    \caption{\textbf{a} Antenna arrays of varying length (scale bar is 1 $\mu$m) and the resulting SRAM map in an unstrained state showcasing the single antenna sensitivity of the setup. \textbf{b} FEM simulation of the slot antenna exhibiting a magnetic dipolar resonance showing the magnetic and electric field distributions at the resonance frequency. The grey arrows indicate the strain direction. \textbf{c} SRAM spectra of a strained antenna array of 7x7 (shown in \textbf{a}). Increasing strain states are indicated with different colours (green to black). The resulting spectra are fitted with a double Lorentzian function. \textbf{d} Strain dependence of the fitted parameters (resonance shift $\Omega_{1,2}-\Omega_0$, amplitude $A_{1,2}$ and line width $\gamma_{1,2}$) of the double Lorentzian function, showcasing the strain sensitivity of the primary eV resonance. }
    \label{fig2}
\end{figure*}

To this extent, we engineer a set of antenna arrays with varying antenna numbers, from a 7x7 array down to a single nanoantenna (Figure~\ref{fig2}a).  This set is fabricated on a template-stripped 100 nm thick gold film on polycarbonate by focused ion beam milling. Polycarbonate is a flexible substrate which allows for a controlled deformation of the metallic thin film and the FIBed antennas by externally applied uniaxial deformation. The antennas' dimensions are chosen such that the resonance falls within the measurement range of the microscope (1.77 eV - 2.55 eV), which corresponds to a length of 100 nm, width of 50 nm and a pitch of 150 nm.  Using a template-stripped film results in ultrasmooth surfaces with increased optical quality and plasmon propagation length~\cite{mcpeak_plasmonic_2015}. For more details on the fabrication, we refer the reader to the Materials and Methods section.
\par
We begin scanning the entire region of the unstrained nanoantennas with the microscope at a single frequency and map the optical anisotropy (Fig.~\ref{fig2}a). The frequency has been matched with the resonance frequency of a single antenna ($\omega=2.39$ eV). Remarkably, the microscope resolves the optical anisotropy down to a single nanoantenna (Fig.~\ref{fig2}a), highlighting the excellent anisotropy sensitivity of the setup. The response of a single nanoantenna, given by the convolution of the point spread function of the antenna and the microscope, enables us to calculate the lower limit to the resolution. The measured resolution is very close to the theoretical resolution confirming that the optical setup is diffraction-limited (see also Supplementary Information).

\par 
As the antennas are milled into the film, they exhibit a magnetic dipole moment instead of the electric dipole moment observed in typical nanorod antennas~\cite{park_optical_2018,hrton_plasmonic_2020}. Figure \ref{fig2}b shows the normalized electric field intensity and magnetic field lines calculated with FEM simulations of the slot antennas, confirming their magnetic dipolar nature.  
\par
To mechanically deform the nanoantennas, the flexible substrate is clamped in a tensile stage and strained until a flat profile is obtained so that the light beam illuminates at normal incidence on the film. The film has an undetermined initial strain state and all subsequent strain measurements are given relative to the initial strain state. Figure~\ref{fig2}d shows the SRAM spectra of an antenna array taken for 5 incremental strain states. Since the antennas are magnetic dipoles, their line shapes can be represented by a Lorentzian (see also Supplementary Information). The fitted Lorentzian line shapes shown in Figure~\ref{fig2}d are in excellent agreement with the experimental spectra. Despite the small deformation, the resonance frequency at 2.39 eV shifts to longer wavelengths with increasing strain up to a shift of $\Delta\Omega_1=-37.7\pm3.52$ meV, corresponding to a strain sensitivity of $\kappa=-20.9\pm1.53$ meV/\%. The shift is also observed to be linear with strain (Figure~\ref{fig2}c), which fits expectations of plasmonic nanorod antennas~\cite{knight_aluminum_2012}.

A closer look on the amplitude and line width of the central peak in Figure~\ref{fig2}d, reveals that while the amplitude is decreasing, the line width is increasing resulting in an overall broadening of the resonance with increasing strain. This is due to the mechanical deformation of the nanoresonators. While elongation of the longitudinal axis of the antennas leads to the reported red-shift of the resonance, the small mismatch in Poisson's ratio between the polycarbonate substrate and the gold film and the induced deformation on the film lead to a non-homogeneous shape change of the resonators. Cracks and considerable surface deformation are observed for strains higher than 2\%, when the resonance was not observable anymore. We also observe a second resonance at 1.98 eV. This resonance does not show a conclusive strain dependence in any of the fitted parameters and is presumably related to the onset of plastic deformation.

Quite interestingly, the strain-induced geometrical change of the antennas can be considered a model system that mimics the strain-induced symmetry breaking in conventional materials. Specifically, the observed shift in the dipolar resonance of the antennas can be seen as an analogue to the energy shift of the dipolar transitions at the high symmetry points of the band structure of conventional materials~\cite{volpi_sensing_2021}. This energy shift is typically described within the framework of deformation potential theory. In the next sections, we measure the optical anisotropy induced by these transitions using SRAM for two conventional material classes: semiconductors and metals.

\subsection{Microscopy of Strained Semiconductors}

When a semiconductor is mechanically deformed, the associated strain induces a change in the electronic band structure, which is most pronounced at critical points associated with interband transitions. This translates to a change in the dielectric permittivity and gives rise to a materials dependent optical signal. Especially epitaxial systems, due to their low roughness, defect density and large grain size, are well suited to study the deformation of the electronic band structure by optical means.
\par  
For small anisotropy in the dielectric function ($\varepsilon_1$, $\varepsilon_2=\varepsilon_1+\Delta\varepsilon$, $\Delta\varepsilon\ll\varepsilon_1$), the optical response can be written as~\cite{weightman_reflection_2005}

\begin{figure*}[ht!]
    \centering
    \includegraphics[width=\linewidth]{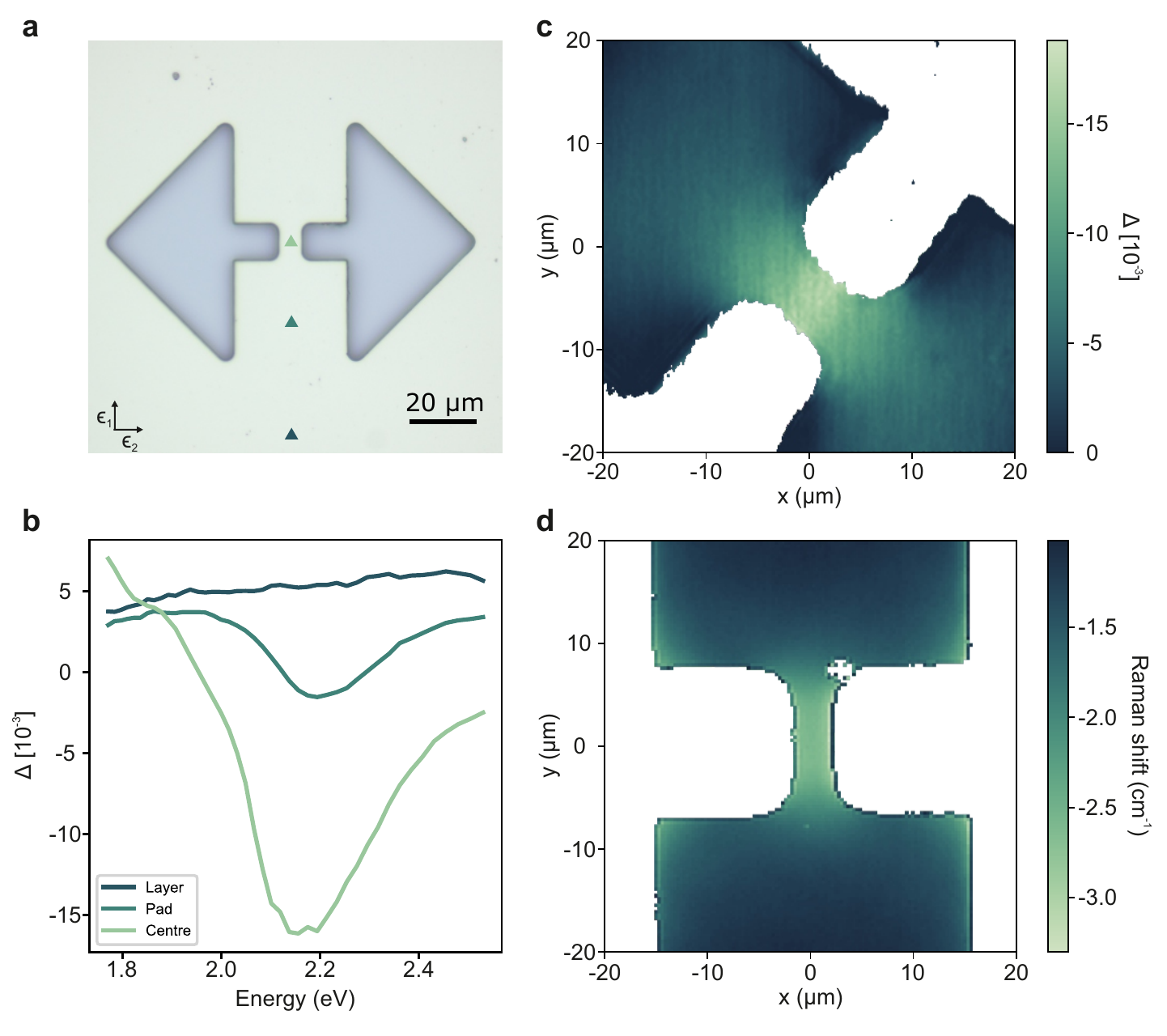}
    \caption{\textbf{a} Optical microscope top view of suspended and highly uniaxially strained germanium microbridges. The colored triangles indicate the position where the $\mu$RAS spectra in \textbf{b} were taken. \textbf{b} Measured spectra for different points of the structure. The highly uniaxial strain of the bridge yields a RAS resonance while the equibiaxial strain of the unreleased germanium layer yields no signal. \textbf{c} $\mu$RAS map at 2.21 eV (resonance frequency of the spectra in \textbf{b}) of the micro-bridges showing the $\epsilon_{xx}-\epsilon_{yy}$ strain distribution. The bridges are oriented at 45$^{\circ}$ to align with the polarisation axis of the PEM and polarizer/analyzer. \textbf{d} Raman shift map of the micro-bridges, proportional to $p\epsilon_{zz}+q(\epsilon_{xx} + \epsilon_{yy})$.}
    \label{fig3}
\end{figure*}
\begin{equation}
    \frac{\Delta r}{r}=\frac{\Delta\varepsilon}{\sqrt{\varepsilon}\left(1-\varepsilon\right)} ,
\end{equation}

where $\Delta\varepsilon$ is the strain induced change in the unperturbed dielectric tensor $\varepsilon$. Generally, the change in the relative permittivity tensor is related to the strain tensor via the elasto-optic tensor, which is defined as

\begin{equation}
    \Delta\varepsilon_{i}^{-1}=p_{ij}\epsilon_{j}\label{defP},
\end{equation}

where $\epsilon_j$ are the components of the strain tensor and symmetry arguments have been used to reduce index notation to $i,j=1,2,...,6$ \cite{nye_physical_1985}. For small strains, equation \ref{defP} can be approximated to

\begin{equation}
    \Delta\varepsilon_{i}=-\varepsilon^2 p_{ij}\epsilon_{j} .
\end{equation}

For a cubic material ($m3m$), such as silicon and germanium, the elasto-optic tensor has three independent components, $p_{11}$, $p_{12}$ and $p_{44}$. In the case were the direction of observation is one of the principal crystal axis, e.g. [001], the difference in dielectric permittivity and reflectance can be written as

\begin{equation}
   \Delta\varepsilon=\Delta\varepsilon_{[100]}-\Delta\varepsilon_{[010]}=-\varepsilon^2\left(p_{11}-p_{12}\right)(\epsilon_1 - \epsilon_2) ,
\end{equation}

\begin{equation}
    \frac{\Delta r}{r}=-\frac{\varepsilon^{3/2}}{1-\varepsilon}\left(p_{11}-p_{12}\right)(\epsilon_1- \epsilon_2) ,\label{theo_ras_vs_strain}
\end{equation}

where $\epsilon_1=\epsilon_{xx}$ and $\epsilon_2=\epsilon_{yy}$. Unlike Raman spectroscopy, the out of plane strain $\epsilon_{zz}$ and off-diagonal components don't have any effect on the optical response when the direction of observation is parallel to one of the principal axes of the crystal.

Here, we make use of suspended germanium micro bridges that achieve high uniaxial tensile loads (Figure \ref{fig3}a) to measure the resonance frequency of the irreducible component of the elasto-optic tensor $\left(p_{11}-p_{12}\right)$. Uniaxial tension is achieved by depositing a thermally biaxially strained germanium layer on a silicon wafer and then using reactive ion etching to partially release the strain, resulting in highly strained microbridges. A detailed explanation of the fabrication method and a thorough characterization of the structure is available elsewhere~\cite{suess_analysis_2013}. 
\par
Figure \ref{fig3}b shows the SRAM spectra taken at different spots of the bridge structure (the measured spots are indicated in Figure \ref{fig3}a). 
The signal is highest at the center of the bridge and completely disappears on the unreleased wafer, as expected for the equibiaxially strained layer. Fitting a Lorentzian lineshape to the peak at the center of the bridge yields a resonance frequency of 2.21 eV, which is in the range of the $E_1$ and $E_1+\Delta$ bandgap and in good agreement with previous measurements of the piezo-optic tensor components of germanium~\cite{etchegoin_piezo-optical_1992}. 
\par
Scanning the entire bridge at the resonance energy enables the determination of the local  differential strain $\epsilon_{xx} - \epsilon_{yy}$. In excellent agreement with experiments and FEM simulations described in earlier works~\cite{suess_analysis_2013}, the differential strain concentrates at the center of the bridge and falls off further away into the pad (Figure \ref{fig3}c). 
\par
To further assess the strain mapping capabilities of our microscope, we compare with an established optical technique for strain mapping: Raman spectroscopy. Figure \ref{fig3}d shows a map of the Raman shift for the same microbridge as in Figure \ref{fig3}c. Details of the experimental procedure can be found in the Materials and Methods section. As can be observed, the Raman shift follows a slightly different distribution than the one measured with SRAM. This discrepancy is due to the different strain linear combinations that these two techniques probe. While SRAM is proportional to $\epsilon_{xx} - \epsilon_{yy}$, the Raman shift is proportional to $p\epsilon_{zz}+q(\epsilon_{xx} + \epsilon_{yy})$~\cite{DeWolf2015}, where p and q are the Raman tensor coefficients.
\par

While both SRAM and Raman microscopy provide a way to infer the mechanical state of the sample, the underlying mechanisms are intrinsically different. SRAM directly probes the anisotropy in the dielectric permittivity tensor while Raman spectroscopy probes the vibrational levels of the crystal lattice that present a change in polarizability. As such, SRAM offers direct insight into the optical properties of the sample and has access to a broader range of materials, e.g. metals, for which a case study is presented in the following section.

\subsection{Microscopy of Strained Metals}

Similar to $m3m$ semiconductors, density functional theory calculations have shown that the strain sensitivity arises from polarization dependent interband transitions~\cite{volpi_sensing_2021}. However, unlike Germanium or Silicon, it is often the case that metals exhibit a polycrystalline microstructure, which can be either textured or random, 
with average grain size significantly smaller than the focal spot of the light beam. In such a scenario, the measured optical signal is an average over all the grain orientations such that the photo-elastic tensor becomes a scalar and equation \ref{defP} becomes a scalar relation.

\begin{figure*}[!htb]
    \centering
    \includegraphics[width=1\linewidth]{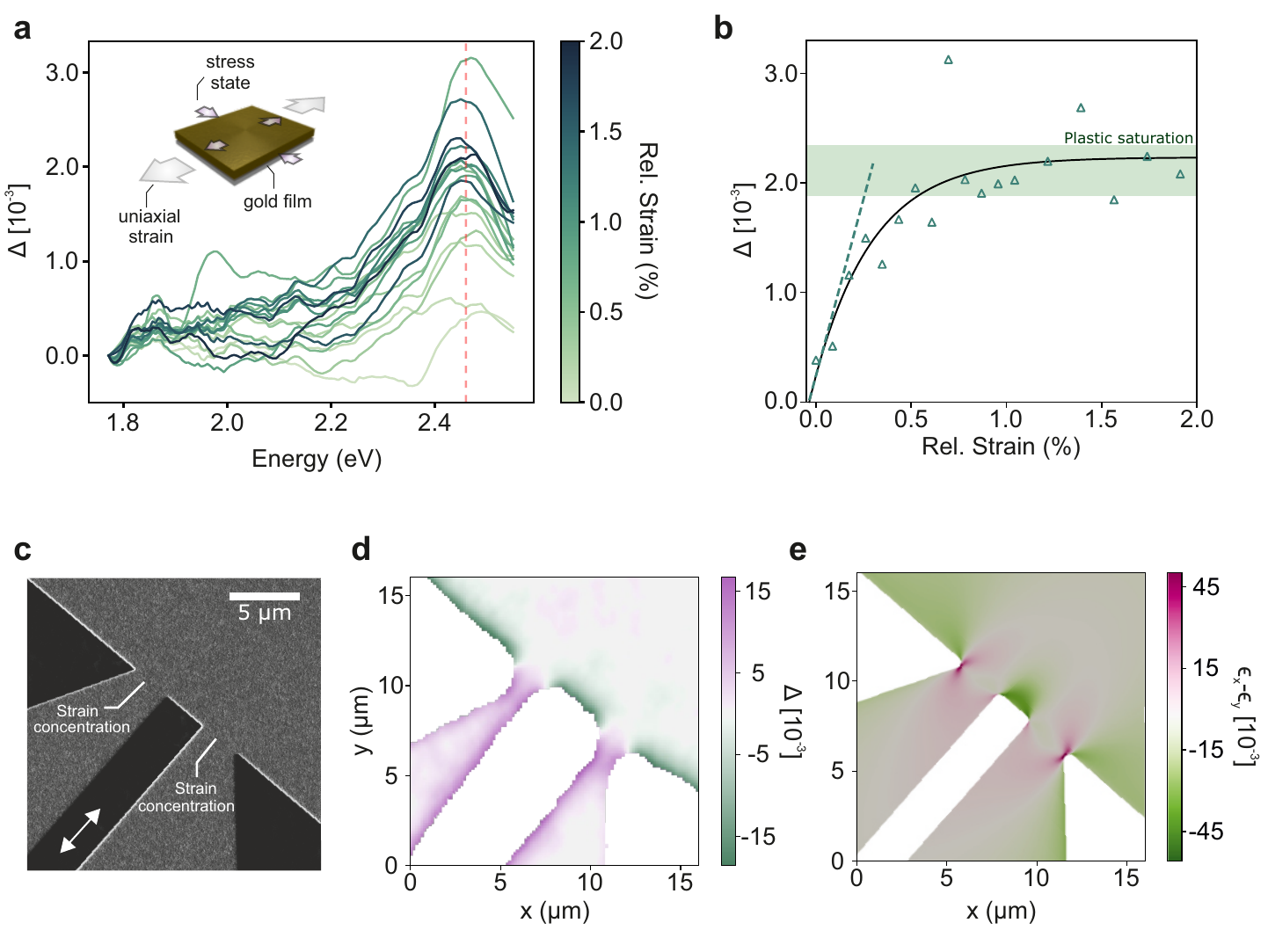}
    \caption{\textbf{a} SRAM spectra of an externally strained gold film deposited on a polyimide substrate (Kapton E) as a function of strain showing a resonance at 2.44 eV. \textbf{b} SRAM signal at 2.44 eV as a function of strain. An exponential function derived from the constitutive law of Voce (equation \ref{eqVoce}) provides the best fit for both the linear elastic regime and the plastic saturation regime (indicated in light green). \textbf{c-e} FIBed double wedge structure and the resulting strain distribution, obtained with both the SRAM setup (\textbf{d}) and FEM simulations (\textbf{e}). The strain direction is indicated with an arrow in \textbf{c} together with the expected zones of strain concentration and relaxation.}
    \label{fig4}
\end{figure*}

To this extend, we apply SRAM to strained gold films. Due to their excellent optical, electrical and mechanical properties, gold films are readily used in flexible electronics. To produce films that better simulate conditions typically found in these flexible devices, we refrained from template-stripping in the fabrication of the films.
\par
Figure~\ref{fig4}a shows a set of SRAM spectra as a function of strain of a 500 nm thick gold film sputtered on polyimide (Kapton E), which presents a Poisson's ratio similar to that of gold. The characteristic peak at $2.44$~eV in all spectra (Figure~\ref{fig4}a) corresponds to the interband transition near the high symmetry point L between states in the 5d and 6sp bands of gold~\cite{guerrisi_splitting_1975,Ngoc2015}. Similar to the case of semiconductors, the amplitude of the resonance peak (Figure \ref{fig4}b) scales linearly with strain in the elastic regime. However, in the case of metals for strains higher than the yield point one has to take into account plastic deformation and its effect on the strain sensitive signal.
\par
In the plastic regime, the reflectance anisotropy signal should eventually saturate as further strain generates dislocations but does not increase the lattice spacing, which is the origin of the photoelastic effect. Instead, further increases in signal after the film yields are due to the creation of anisotropic defects that introduce a global offset to the spectra in the bandwidth of the SRAM~\cite{calvo_reflectance_2022,blackford_ras_2005}. These contributions are low in the elastic regime but are significant in the plastic regime, specially for sputtered thin films that have increased roughness compared to template stripped films or epitaxial semiconductors. To uncover the signal stemming from strain in the plastic regime, we introduce a correction term at 1.77 eV, far from resonance where the spectra should be equal.

Using an exponential function analog to the constitutive law of Voce to fit the plastic saturation (Figure \ref{fig4}b) \cite{Wyss2017}, we can determine the average photoelastic constant of the thin gold film to be $\overline{P}=(1.83-3.03i)\num{e-3}$. More information is given in the Supplementary Information.

\par
In order to showcase the capabilities of SRAM to map strain in metals, we test more complex metallic structures with a spatially anisotropic strain distribution. We fabricate a double wedge microstructure using FIB milling that introduces a strain distribution when externally applying uniaxial tension (Figure \ref{fig4}c). Such a structure should generate a strain concentration at the wedge gaps and strain relaxation close to the edges perpendicular to the strain direction (see Figure \ref{fig4}c). Applying a strain lower than the elastic limit (strain direction indicated with a double arrow in Figure \ref{fig4}c) ensures that plastic saturation is not reached so that the SRAM map is proportional to the strain distribution. Figure \ref{fig4}d shows the SRAM map at 2.44 eV, where the optical response has been suppressed at the FIBed structure. This is done to filter artefacts generated by the metal edges, which break the symmetry of the focal spot. As can be seen, the SRAM map shows the expected concentration and relaxation of strain around the double wedge structure.

To further validate our experimental results, we carried out FEM simulations of the strain distribution generated by the FIBed structure (Figure \ref{fig4}e) and find good agreement with the measured SRAM map. The regions with strain concentration/relaxation are reproduced by the measured SRAM strain distribution. The SRAM data does, however, show that the strain concentration/relaxation induced by the double wedge structure decays faster compared to FEM simulations. This is explained by the polyimide film being two orders of magnitude thicker than the metal film, thus dominating the mechanics of the system. Applying a shear lag model to calculate the strain transfer from the substrate to the film we find the stress transfer length to approximately be 4 $\mu m$, in agreement with the measured SRAM maps. More details can be found in the Supplementary Information.

The higher roughness (sub-wavelength) and polycrystallinity of sputtered metals in comparison to typical semiconductor surfaces translates to an increased noise level in the SRAM signal and slighthly reduces the high strain sensitivity of the technique. However, the noise is not structured and is, therefore, mitigated by the spatial mapping of the setup. As seen by the good match between the SRAM maps and FEM simulations the noise does not impede an accurate characterization of the strain distribution.

\section{Discussion}

Non-destructive mechanical characterization has become essential for an improved understanding and  design of complex multi-material devices, such as flexible electronics, that comprise demanding mechanical deformation conditions. The scanning reflectance anisotropy microscope (SRAM) presented herein provides a versatile stand-alone platform for high-resolution strain mapping and strain imaging of semiconductors, metasurfaces and metals. To the best of our knowledge, SRAM is currently the only optical technique able to map strain in metals without strain markers. Especially, the use of a supercontinuum source instead of a spectrometer makes SRAM a versatile, low-footprint and non-invasive technique, that can be readily extended to access materials, e.g. with fundamental transitions in the near- to mid-infrared, potentially enabling characterization of soft materials. 
\par

In addition to its broad applicability, we show that the microscopes reaches a 500-nm-level spatial resolution, a smaller than $1^\circ$ phase sensitivity and a strain sensitivity $\kappa=-20.9$ meV/\% . As SRAM is mainly limited by noise-contributions stemming from roughness or topography, we expect that learning algorithms for data treatment can further improve the strain sensitivity of the technique.

\par

SRAM can also be used in conjunction with other strain sensing techniques. For example, analyzing complex strain states using Raman spectroscopy generally requires the usage of FEM simulations to decouple the different strain components. However, for plane stress, the combined use of SRAM ($\propto\epsilon_{xx} - \epsilon_{yy}$) and Raman ($\propto\epsilon_{xx} + \epsilon_{yy}$) would allow to decouple $\epsilon_{xx}$ and $\epsilon_{yy}$ without the assistance of FEM simulations.

\par
As SRAM locally measures the in-plane ellipsometric parameters of a material, we envisage SRAM to contribute also to the design and characterization of complex metamaterials, such as metalenses or chiral metasurfaces, particularly when strain is used as an external input for parameter tuning.

\section{Materials and Methods}

\subsection{Microscopy setup}
The SRAM setup (Figure \ref{fig1}a) is a phase modulated microscope consisting of a beamsplitter (BSW10R, Thorlabs Inc.) and an objective lens (LMPLFLN50X, Olympus). The light source is a supercontinuum laser (EXR20, NKT Photonics) that is monochromated with an acusto-optic modulator (SuperK SELECT, NKT Photonics) for a spectral range from 485 nm to 700 nm with a bandwidth of 2 nm. The output is coupled into a fiber and collimated with an off-axis parabolic mirror (RC04APC-F01, Thorlabs Inc.). The polarization optics of the setup are comprised of a photoelastic modulator (PEM100, Hinds Instruments Inc.) and two rochon prisms (A-BBO, Edmund Optics Ltd.) that act as polarisers. The reflected light is then focused onto a photomultiplier tube (R10699, Hammamatsu) and a lock-in amplifier (MFLI500, Zurich Instruments Ltd.) locked onto the resonant frequency of the PEM used to measure the amplitude and phase of the ac components of the signal. Both SRAM spectra, maps and hypermaps are taken with a phase locked loop time constant of $t_c=0.05$ s, and an integration time of 1 s per data point and scans have a pixel size of 200 nm. To compensate for focusing artefacts, all spectra are calibrated against a measurement performed on a (100) silicon wafer (Figure S1).

\subsection{Gold thin films}

Gold (99.99\% MaTeck GmbH) films were deposited with a magnetron sputtering machine (PVD Products Inc.) with a chamber pressure of $\num{5e-7}$ Torr, argon plasma pressure of 3 mTorr and 200 W magnetron power onto two different substrates. A 500 nm film was deposited onto a 50 $\mu$m thick polyimide film (Kapton E, DuPont de Nemours, Inc.) for the double wedge experiments. The Kapton E films were cleaned by ultrasonicating for 10 min in aceton, ethanol, and isopropanol, rinsing with isopropanol and drying overnight in vacuum. For the nanoantennas experiments, 100 nm gold films were deposited on a silicon nitride wafer and then template stripped with a polycarbonate film to reduce surface roughness. The template stripping was carried out in a hot-press transfer process with polycarbonate pellets (Makrolon, Bayer AG) at 240~$^{\circ}$C and 20~kN with the films getting cooled under pressure to prevent wrapping~\cite{Deshmukh2018}. Both 500 nm and 100 nm thick films were then cut in thin stripes of 3 mm by 25 mm to homogenise the strain field and reduce necking and then the wedge microstructures and the nanoresonators were milled with focused ion beam (NVision 40, Carl Zeiss AG). The tensile stage used for the mechanical experiments is actuated with a micrometer screw with a graduation of 10 $\mu$m per division (150-801ME, Thorlabs Inc.). As the film has a gauge length of 25 mm, that translates to a strain resolution of $\pm$0.02\%.

\subsection{Raman spectroscopy}
The Raman scans on the Germanium micro-bridges were carried out using a commmercial Raman microscopy setup (LabRAM HR Evolution UV-VIS-NIR, Horiba Ltd.) with a 1800 gr/nm and 500 nm blaze grating and a 532 nm single wavelength laser (1500 mW Nd:Yag, Cobolt Samba). The same objective as in the SRAM setup (LMPLFLN50X, Olympus) was used with a single accumulation of 1 s integration time per point. The obtained spectra were fitted with a Lorentzian function to calculate the Raman shift from the unstrained c-Ge peak at 300.7 cm$^{-1}$.

\subsection{Data Availability}
All data needed to evaluate the conclusions in the paper are presented in the main text and/or the Supplementary Information. Additional data are available from the corresponding author upon reasonable request.

\section*{acknowledgement}
The authors acknowledge the infrastructure
of the microscopy platform of ETH Zurich (ScopeM) for the Raman spectroscopy measurements. The authors thank Nerea Abando for her help and support in the earlier stages of this study. This project has been funded by the Swiss National Science Foundation under Grant No. 172717.

\bibliography{main}

\appendix
\onecolumngrid
\pagebreak
\section*{Supporting Information}
\renewcommand\thefigure{S\arabic{figure}}    
\setcounter{figure}{0}
\renewcommand\theequation{S\arabic{equation}}  
\setcounter{equation}{0}

\subsection{Microscope calibration}
\begin{figure}[ht!]
    \centering
    \includegraphics[width=0.5\linewidth]{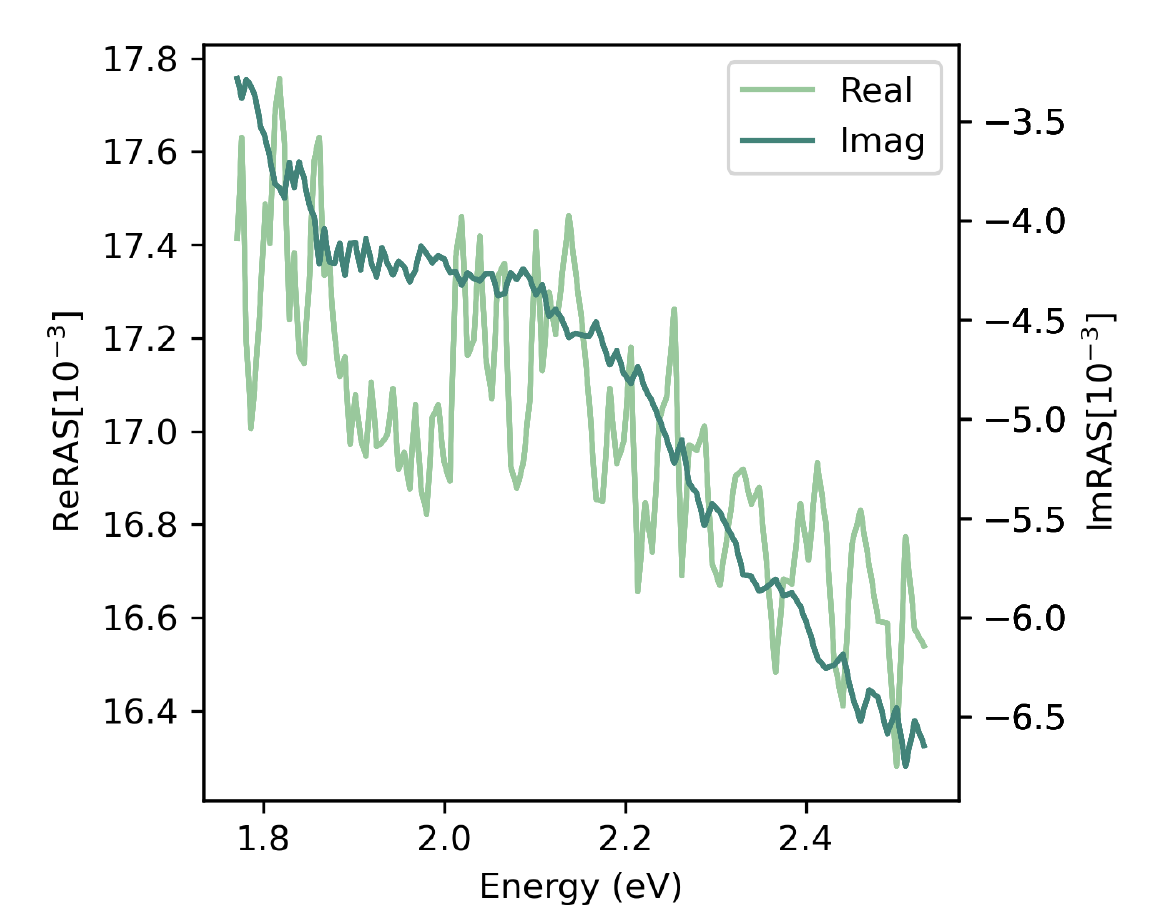}
    \caption{SRAM spectrum of a silicon (100) wafer used for calibration of the setup.}
    \label{figSup2}
\end{figure}

The angular dependence of the Fresnel coefficients comes into play when considering using an objective lens. Nonetheless, Koopmans \textit{et al.}\cite{koopmans_microscopic_1998} and Shen \textit{et al.}\cite{shen_using_2017} showed that the focusing artefacts are compensated by the $x^2 - y^2$ symmetry of the system with good objective alignment. Even though the symmetry of the system compensates the contribution of the angular dependence, in practice small misalignments introduce artefacts in the detected signal. We compensate these artefacts with a calibration measurement (see Figure S \ref{figSup2}) on a silicon (100) wafer, which has a four-fold symmetry and should yield no signal. Similar to ellipsometry, the measurements are normalized by the total reflectance and do not need to be calibrated against the light source spectrum.

\subsection{Single nanoantenna scan}
\begin{figure}[ht!]
    \centering
    \includegraphics[width=\linewidth]{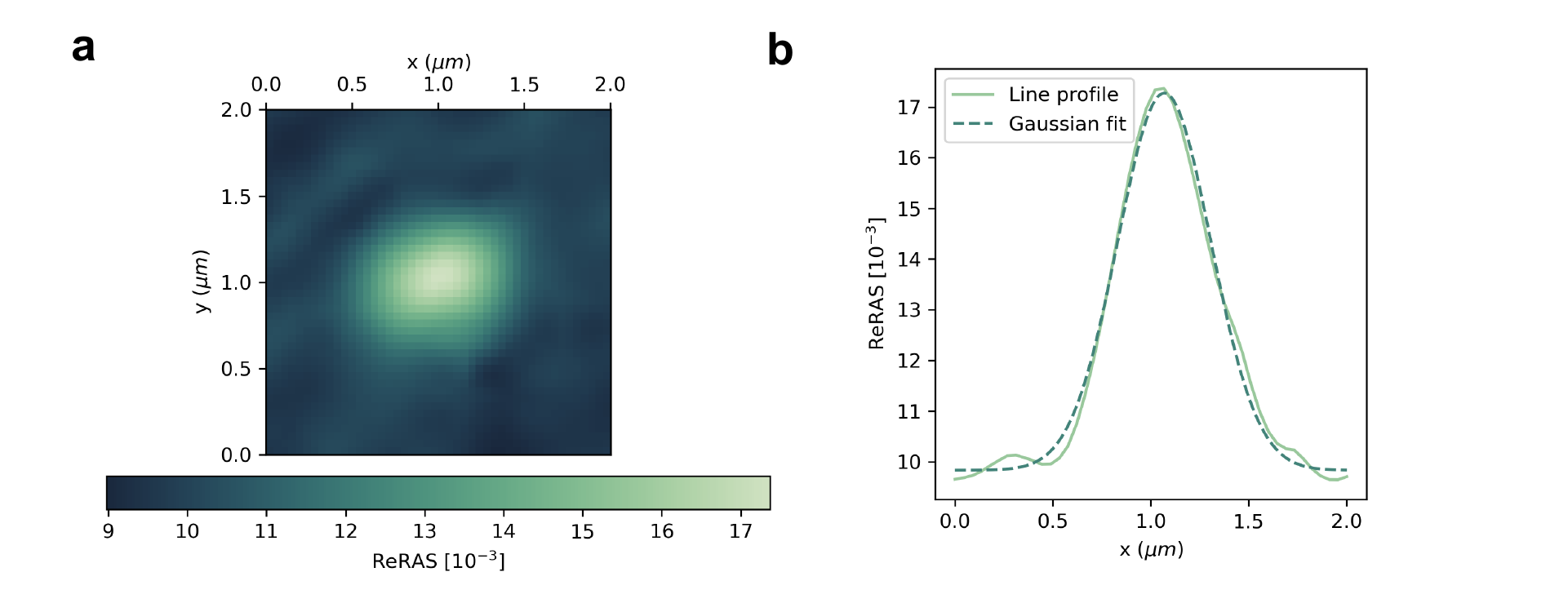}
    \caption{\textbf{a} Single wavelength scan of a single nanoantenna at 2.39 eV. The resulting signal is a convolution of the point spread function of the nanoantenna and the point spread function of the focused light. Since the nanoantenna could be interpreted as a point like structure, the signal is close to the point spread focused beam. \textbf{b} Line profile of the scan in \textbf{a} fitted with a gaussian function. The full width half maximum of the fitted gaussian is 560 nm.}
    \label{figSup1}
\end{figure}

\subsection{Fit of nanoantenna resonance}

Since the nanoantennas present a dipolar resonance, we fit the lineshape with a Lorentzian function

\begin{equation}
    \Psi\left(\omega\right)=\frac{A_1}{\pi}\frac{\gamma_1}{\left(\omega-\Omega_{1}\right)^2+\gamma_1^2}+\frac{A_2}{\pi}\frac{\gamma_2}{\left(\omega-\Omega_{2}\right)^2+\gamma_2^2} ,\label{DoubleL}
\end{equation}

where the set of parameters ($A_1$,$\gamma_1$,$\Omega_1$) --- amplitude, linewidth and resonance frequency, all expressed in units of energy --- describes the resonance at 2.39 eV and the set ($A_2$,$\gamma_2$,$\Omega_2$) describes the resonance at 1.98 eV.

\subsection{Fit of SRAM-strain relation for metallic thin films}

The constitutive law of Voce describes the stress-strain relation in the presence of strain hardening. We use an analogue to this equation, already introduced in reference \cite{Wyss2017}, to fit the SRAM-strain relation:

\begin{equation}
\Delta r/r(\varepsilon)=\Delta r_\infty - (\Delta r_\infty - \Delta r_0)e^{-\frac{W}{\Delta r_\infty - \Delta r_0}\varepsilon},\label{eqVoce}   
\end{equation}
\par
where $\Delta r_0$, $\Delta r_\infty$ and $W$ work as a reflectance anisotropy analogues to the stress parameters in the Voce equation. $\Delta r_0$ is the anisotropy signal at the initial strain state of the film, $\Delta r_\infty$ is the signal at the plastic saturation plateau and $W$ is the initial slope. $W$ represents the constant of proportionality between the reflectance anisotropy signal and elastic strain and, therefore, can be related to the photoelastic tensor of gold.

\subsection{Edge induced strain relaxation}
In a system comprised by a metallic thin film and a substantially thicker substrate (in this case 100 times greater), the substrate dominates the mechanical response of the system. As such, the stress experienced by the metallic thin film is transferred from the substrate. The transfer rate depends the ratio of mechanical properties between the film and the substrate. This is effect is easier seen in regions near film edges where the free surface is relaxed and presents 0 stress while at a sufficient distance away from the edge the film reaches the stress level transferred by the substrate. The stress curve as a function of distance from the free surface is exponential and is characterized by length parameter dependent on the thin film's thickness and the elastic mechanical properties of the film and substrate. We employ the following linear shear lag model to calculate the characteristic length of our gold-polyimide system \cite{xia2000crack}

\begin{equation}
    l=\frac{\pi}{2}g\left(\alpha,\beta\right)h ,
\end{equation}

where $l$ is the reference length, $h$ the film thickness and $g\left(\alpha,\beta\right)$ is a function of the Dundurs parameters $\alpha$ and $\beta$ representing the mismatch in elastic mechanical properties between substrate and film \cite{beuth1992cracking}. Using the values $h=500$ nm and $g=5.17$ (for $\alpha=0.878$ and $\beta=0.250$) we obtain the reference length $l\simeq4$ $\mu$m, which is in very good agreement with the strain decay observed in the SRAM map.


\end{document}